\input ./style/arxiv-general.cfg
\documentclass[aoas,MSNbibl,nameyear,dvips]{arximspdf}
\makeatletter
   \@ifpackageloaded{graphicx}{}{\usepackage{graphicx}}
\makeatother
\doi{10.1214/15-AOAS854}% Updated by VTEXPTS2LaTeX.exe, 21.08.2015 12:12
\volume{9}
\issue{3}
\pubyear{2015}
\firstpage{1643}
\lastpage{1670}
\docsubty{FLA}

\makeatletter
\newcommand{\cal}{\mathcal}
\renewcommand{\epsilon}{\varepsilon}
\makeatother

\begin{document}
\begin{frontmatter}

\title{The Gibbs-plaid biclustering model}
\runtitle{Gibbs-plaid biclustering}
\thankstext{TT1}{Supported in part by NSERC Grant 327689-06.}
\thankstext{TT2}{Supported in part by the funding from
ANR-12-RPIB-0012-03 (OncoVaccine),
ANR-10-BINF-03-02 (BIPBIP), CNRS, INSERM and the Universit\'e de Strasbourg.}

\begin{aug}
% Corresponding author: Alejandro Murua - murua@DMS.UMontreal.CA% Updated by VTEXPTS2LaTeX.exe, 24.08.2015 09:56
%Updated by VTEXPTS2LaTeX.exe, 21.08.2015 12:12
\author[A]{\fnms{Thierry}~\snm{Chekouo}\thanksref{TT1}\ead[label=e1]{Tchekouo@mdanderson.org}},
\author[B]{\fnms{Alejandro}~\snm{Murua}\thanksref{TT1}\corref{}\ead[label=e2]{murua@dms.umontreal.ca}}
\and\\
\author[C]{\fnms{Wolfgang}~\snm{Raffelsberger}\thanksref{TT2}\ead[label=e3]{w.raffelsberger@unistra.fr}}
\runauthor{T. Chekouo, A. Murua and W. Raffelsberger}
\affiliation{University of Texas MD Anderson Cancer Center, Universit\'e de
Montr\'eal, and
ICube/LBGI and IGBMC (CNRS, INSERM, Universit\'e de
Strasbourg)}
%\dedicated{}
\address[A]{T. Chekouo\\Department of Biostatistics \\
University of Texas\\
\quad  MD Anderson Cancer Center\\
1400 Pressler Street, Unit 1411 \\
Houston, Texas 77030-3722\\ USA\\\printead{e1}}
\address[B]{A. Murua\\D\'epartement de math\'ematiques\\
\quad  et de statistique \\
Universit\'e de Montr\'eal\\
CP 6128, succ. centre-ville \\
Montr\'eal, Qu\'ebec H3C 3J7\\ Canada\\\printead{e2}}
\address[C]{W. Raffelsberger\\LBGI, ICube Laboratory and FMTS\\
University of Strasbourg and CNRS, Facult\'e de M\'edecine\\
67085 Strasbourg\\ France\\
and\\ IGBMC (CNRS, INSERM, Universit\'e de Strasbourg)\\
1 rue Laurent Fries, 67404 Illkirch\\ France\\\printead{e3}}
\end{aug}

% HISTORY:
%
\received{\smonth{1} \syear{2014}}% Updated by VTEXPTS2LaTeX.exe,
%21.08.2015 12:12
%
\revised{\smonth{3} \syear{2015}}% Updated by VTEXPTS2LaTeX.exe,
%21.08.2015 12:12

% ABSTRACT
%
\begin{abstract}
We propose and develop a Bayesian plaid model for biclustering that
accounts for the prior
dependency between genes (and/or conditions) through a stochastic
relational graph.
This work is motivated by the need for improved understanding of the
molecular mechanisms of human diseases for
which effective drugs are lacking, and based on the extensive raw data
available through gene expression profiling.
We model the prior dependency information from biological knowledge
gathered from gene ontologies. Our model, the Gibbs-plaid model,
assumes that the relational graph is governed by a Gibbs random field.
To estimate the posterior distribution
of the bicluster membership labels, we develop a stochastic algorithm
that is partly based on the
Wang--Landau flat-histogram algorithm. We apply our method to a gene
expression database created from
the study of retinal detachment, with the aim of confirming known or
finding novel subnetworks of proteins associated with this disorder.
\end{abstract}

% KEYWORDS
\end{frontmatter}

%s1 #&#
\section{Introduction}\label{secintro}
\setcounter{footnote}{2}
DNA microarray and sequencing
technologies allow investigators to measure the transcription\footnote{%
Transcription is the copying of DNA segments into RNA.
}
levels of a large numbers of genes within several diverse experimental
conditions (or experimental samples)
[\citet{Sara-Oliveira-2004,Tanay-2005}].
The experimental conditions may correspond to either different time
points, different environmental samples, or different individuals or
tissues. The data resulting from these technologies are usually
referred to as \textit{gene expression data}.

A gene expression data set may be seen as a data matrix, with rows and
columns respectively corresponding to genes and experimental
conditions. Each cell of this matrix represents the expression level of
a gene under a biological condition. The analysis of gene expression
data usually implies the search for groups of co-regulated genes, that
is, groups of genes that exhibit similar expression patterns.
Inversely, the analysis may seek samples or conditions (e.g., patients)
with similar expression profiles. These may indicate the same
attribute, such as a common type or state of a particular disease.
Vast amounts of gene expression data from numerous experiments are
available for detailed analysis through public repositories such as the
Gene Expression Omnibus (GEO) [\cite{GEO-2002}] at the National Center
for Biotechnology Information.

In general, unveiling the hidden structure in gene expression data
requires the use of
exploratory analytical methods such as \textit{clustering}.
Cluster analysis has been used successfully to analyze a wide variety of
transcriptomes\footnote{A transcriptome is the collection within a cell
of all the messenger RNA, which transcribes the genetic information for
protein synthesis.}
[e.g., see the review by
\cite{Kerr-et-al-2008}].
As all major biological functions are built on the synergistic
interplay of multiple proteins (the role of genes is to produce proteins),
clustering similar gene expression patterns into distinct groups
corresponds with the belief that different genes that are regulated and
co-expressed
at the same time and in similar locations are likely to contribute to
the same biological functions.
Classical clustering analysis (e.g., the popular $K$-means algorithm
[\cite{Ward-1963}]) associates a given gene with only one cluster.
Moreover, all genes in a given cluster must show similar co-regulation
patterns across all experimental conditions.
These are very stringent conditions for gene expression, as
a given protein (the product of a gene) may have the capacity to
regulate several different biochemical reactions.
In fact, many proteins intervene in a number
of different biological processes or biochemical functions, as
documented in the Gene Ontology (GO) project [\cite{GO-2000}], a major
bioinformatic initiative to unify the representation of gene and gene
product attributes across all species.
The GO project provides an ontology of controlled vocabularies that
describes gene products in terms of their associated biological
processes, cellular components and molecular functions in a
species-independent manner.
Classical clustering of genes (or conditions) cannot assign a gene (or
a condition) to several different clusters.
The approach of biclustering better accommodates the multi-functional
character of genes
across subsets of experimental
conditions. Biclustering is the simultaneous clustering of genes (rows)
and conditions (columns).
In biclustering, a given gene may be associated simultaneously with
several different clusters, which may describe distinct biological
processes that are run by a cell at a given time and which use a given
set of proteins.
\cite{Hartigan-1972} seems to be the first to have applied a clustering
method to simultaneously cluster rows and columns. He introduced the
so-called {\em direct clustering} algorithm, a partition-based
algorithm that allows
for the division of data into submatrices (biclusters).

We apply our methods to the analysis of gene expression data associated
with retinal detachment (RD), a disorder of the eye that typically
leads to permanent vision loss. RD occurs when the sensory layer of the
retina (a thin tissue lining the back of the eye) pulls away from the
pigmented layer of the retina. This results from atrophy or tearing of
the retina secondary to a systemic disease such as diabetes or from
injury or other disturbances of the eye that allow fluids to enter the
space between the sensory and pigmented retinal layers [\cite
{Franklin-2002}]. Surgical intervention to remove the detached parts of
the retina is the current standard of care to prevent further
progression of the disorder. If not treated properly, the entire retina
will progressively detach, leading to complete blindness. Better
knowledge of the molecular mechanisms involved in the progression of RD
is of great interest in order to develop novel drugs to stop or slow
the detachment process, either as a substitute for surgical
intervention or to use in combination with surgical intervention.

Molecular events that occur during the
progression of RD were studied via transcription profiling
[\cite{Delyfer-2011}].
Briefly, 19 retinal biopsies from patients with RD were compared to 19 normal
retinal samples using Affymetrix microarrays. These arrays covered the
human genome, with 54,000 probe-sets.\footnote{A {\em probe} is a
general term for a ``piece of DNA or RNA'' corresponding to a gene or
genetic sequence of interest.
Groups of probes are combined into probe-sets, and multiple probe-sets
may exist for a single gene.
Here, we use the terms {\em probe-set} and {\em gene} interchangeably.}
The microarray data are publicly available at the National Center for
Biotechnology Information
GEO website [\cite{GEO-2002}] as GSE28133.

Transcriptional changes in photoreceptor cells in the retina are the
primary target for drug development. In an initial analysis of the
retinal transcriptome, \cite{Delyfer-2011} used $t$-tests (as is normally
done by bioinformatic labs) to compare normal versus RD samples. In
that analysis, the RD inflammatory response dominated any other
transcriptional changes [\cite{Delyfer-2011}]. Inflammation typically
represents a secondary event that follows the initial stimulus that
caused the first tissue detachment. Unfortunately, the more subtle
transcriptional changes in the photoreceptor cells related to the RD
disorder were not well detected.

In the study by \cite{Delyfer-2011}, mutual information techniques
indicated that changes existed in the RD transcription profile other
than those associated with inflammation,
and that they may be a starting point for studying transcriptomic
changes associated with the photoreceptor cells in RD.
However, the mutual information procedure applied in that study
involves iterative optimization of the results and appears to be rather
difficult to automate.
In this work, we analyze the RD data with biclustering techniques. We
choose biclustering techniques because traditional clustering
approaches are not well suited for the analysis of proteins. Some
proteins assume multiple functions and/or work as hubs that mediate,
link or simultaneously synchronize
multiple biological processes (such as the protein TP53 or STAT1
[\cite{Jolliffe-et-al-2013,Stark-et-al-2012}]). Such characteristics of
proteins make it very challenging to use traditional approaches for
clustering the protein interaction networks. Biclustering is well
adapted to this aim. In RD, anti-inflammatory reactions try to stop or
slow the further advancement of the detachment, while apoptotic (i.e.,
cell death) mechanisms degrade the parts of the retina that have been
detached too long and where the fragile photoreceptor cells have
already started to die.
As the retina is composed of three layers with more than eight
different cell types [\cite{Wassle-2004}], studying the behavior of
photoreceptor cells is complex, and biclustering represents a major
advantage when needing to account for the multiple overlapping
functional responses that occur during RD.

Good surveys of existing biclustering algorithms are available
[\cite{Sara-Oliveira-2004,Tanay-2005,Prelic-et-al-2006}].
Cheng and Church's algorithm [\cite{Cheng-Church-2000}] and the plaid model
[\cite{Lazzeroni-Owen-2002}]
are two of the most popular biclustering methods.
It appears that \cite{Cheng-Church-2000} were the first authors to
propose the term biclustering for the analysis of microarray data.
Their algorithm consists of a greedy iterative search that aims to minimize
the mean squared residual error. \cite{Lazzeroni-Owen-2002} proposed
the popular plaid model. They assumed that the expectation of each cell
in the data matrix is formed with the contribution (sum) of different
biclusters. Others have generalized the plaid model into a Bayesian framework
[\cite{Gu-Liu-2008,Caldas-2008,Zhang2010,Chekouo-Murua-2012}].

From our review of the literature, it is apparent that most models used
for biclustering do not take into account application-specific prior
information about genes or conditions and pairwise interactions between
genes or conditions.
In this work, we propose a model that accounts for this information.
We adopt a Gaussian plaid model as the model that describes the
biclustering structure of the data matrix. In addition, we incorporate
prior information on the dependency between genes and between
conditions through dedicated relational graphs, one for the genes and
another for the conditions. These graphs are conveniently described by
auto-logistic models
[\citeauthor{Besag1974} (\citeyear{Besag1974,Besag2001}), \cite{Winkler2003}]
for genes and conditions.
The distributions are pairwise-interaction
Gibbs random fields for dependent binary data. They can
be interpreted as generalizations of the finite-lattice Ising model
[\cite{Ising1925}], which is a popular two-state discrete mathematical
model for assessing ferromagnetism in statistical mechanics.
We will refer to our overall model as the {\em Gibbs-plaid}
biclustering model.

Our prior is elicited from similarities obtained from the GO
annotations.\footnote{%
GO typically has two components: (A) the ontologies themselves, which
are the defined \textit{terms} and the structured relationships between
them (GO ontology); and (B) the associations between gene products and
the terms (GO annotations).
A gene product is a biochemical material (RNA or protein) that results
from the expression of a gene.
Both the GO ontologies and GO annotations are provided by the GO
project in three domains: (i) a cellular component, which refers to the
place in the cell where a gene product is active; (ii) a biological
process, which refers to a biological objective to which the gene or
gene product contributes; and (iii) a molecular function, which refers
to the elemental activities of a gene product at the molecular level.}
An $r$-nearest-neighbor graph over the genes is built from these
similarities. A key parameter of the auto-logistic prior is the
so-called temperature parameter $T$ (due to its analogy with the
physical process of tempering).
The normalizing constant of this prior is, in general, unknown and
intractable. However, for computational purposes, this constant is
needed to implement a stochastic algorithm that aims to estimate the
posterior distribution of the genes' bicluster memberships when $T$ is unknown.
This means that the usual MCMC Metropolis--Hastings procedure is not
applicable to our model. Instead, we adopt a hybrid procedure that
mixes the Metropolis--Hastings sampler with a variant of the Wang--Landau
algorithm
[\cite{WangLandau2001,AtchadeLiu2010,Murua-Wicker-2012}]. The
convergence of the proposed algorithm to the posterior distribution of
the bicluster membership is guaranteed by the work of \cite{AtchadeLiu2010}.

We note that some earlier attempts to incorporate gene dependency
information are available in the literature, but they were carried out
within the context of clustering (as opposed to biclustering) and
variable selection.
\cite{VannucciStingo2011} provide a nice review. \cite{Stingo2011}
proposed a Bayesian model that incorporates information on pathways and
gene networks in the analysis of DNA microarray data.
They assumed a Markov random field prior to capture the gene--gene
interaction network. The neighborhood between the genes uses the
pathway structure from the Kyoto Encyclopedia of Genes and Genomes
(KEGG) database
[\cite{kanehisa2000}]. \cite{Hang2009} and \cite{Vignes2009}
have also used biological information to perform a clustering analysis
of gene expression data.
\cite{Park2007} incorporated GO annotations
to predict survival time and time to metastasis for breast cancer
patients using gene expression data
as predictor variables.
The Potts model has also been used for clustering analysis of gene
expression data
[\cite{Murua-et-al-2008,Getz2000}]. However, in these approaches, the
Potts model [\cite{Sokal}] was used directly as a nonparametric model
for clustering [\cite{Blatt1996}], and not as a prior that accounts for
the gene--gene interaction on another clustering model.

This paper is organized as follows. Section~\ref{sec:model} introduces
the proposed Gibbs-plaid model for biclustering. Section~\ref
{sec:posterior} describes the stochastic algorithm used to estimate the
posterior distribution of the model parameters. This includes the
combination of the Wang--Landau algorithm with the Metropolis--Hastings
sampler. Section~\ref{sec:experiments} shows the results of a
simulation carried out to study the performance of the Gibbs-plaid
model and of the model selection criteria used to determine the number
of biclusters present in a data set.
Section~\ref{sec:applications:RD} deals with the application of our
methodology to the RD data. The supplementary material
[\cite{Chekouo-et-al-supp-2015}] provides more complete results of our
application to the RD data and a high-resolution image
of Figure~\ref{fig:Bicluster4:network}.

%s2 #&#
\section{The model}\label{sec:model}

Let $p$ be the number of genes, and $q$ be the number of experimental
conditions.
Let $Y_{ij}$ denote the logarithm of the expression level of gene $i$
under condition $j$
($i=1,\ldots,p$, $j=1,\ldots,q$). Even though we actually work with the
logarithm of the expression level, we refer to $Y_{ij}$ as the
expression level.
Let $K$ be the number of biclusters.
For all $i$ in the set of genes, $j$ in the set of conditions, and
$k=1, \ldots,K$, we define
the binary variables $\rho_{ik}$ and $\kappa_{jk}$ as taking values in
$\{0,1\}$, so that
$\rho_{ik}=1$ if and only if gene $i$ belongs to bicluster $k$,
and $\kappa_{jk}=1$ if and only if condition $j$ belongs to bicluster $k$.
The symbols $\rho_{i}$ and $\rho$ denote the $K$-dimensional vector of
components
$\{\rho_{ik}\}_{k=1}^{K}$ and the $pK$-dimensional vector comprising
all the vectors $\rho_{i}$, $i=1, \ldots,p$,
respectively. The symbols $\kappa_{j}$ and $\kappa$ are similarly
defined for the conditions.

%s2.1 #&#
\subsection{The plaid model}
\label{sec:plaid:model}

Let $\Theta$ denote the set of parameters of the model, which are made
explicit hereafter.
In the plaid model,
$Y_{ij} = \mu_{ij}(\rho,\kappa,\Theta) + \epsilon_{ij}$,\vspace*{1pt} where $\epsilon_{ij}$
is a zero-mean error term
and
$\mu_{ij}(\rho,\kappa,\Theta) = \mu_0 + \sum_{k=1}^K ( \mu_k + \alpha
_{ik} + \beta_{jk}) \rho_{ik}\kappa_{jk}$,
where $\mu_0$ denotes the overall data mean, and
$\alpha_k=\{\alpha_{ik}, i=1,\ldots,p\}$ and $\beta_k=\{\beta_{jk},
j=1,\ldots,q\}$ are the gene and condition effects
associated with bicluster $k$, measured as deviations from the
bicluster mean $\mu_0 + \mu_k$, $k=1,\ldots,K$.
Hereafter, we denote by $\mu$ the vector of means $(\mu_1, \mu_2, \ldots,\break \mu_K)$.
The model parameters are given by $\Theta=(\mu_{0},\mu, \alpha,\beta)$.
The most common distribution for the error term is a Normal$(0, \sigma
^2)$ distribution
[\cite{Gu-Liu-2008,Caldas-2008,Zhang2010}]. This is the model we adopt here.
In the context of gene expression data, the plaid model is a model for
the logarithm of the gene expression levels.
In the presence of extreme observations, a more robust model may be
more appropriate, such as one with Student-$t$ distributed errors.
Although some researchers have modeled the log-expression with
more complex distributions such as Gamma or double exponential
distributions [\cite{Purdom-Holmes-2005,Newton-et-al-2001}],
the associated achievement of any gains within the context of
biclustering is arguable.
In fact, the simulation study in \cite{Chekouo-Murua-2012} showed that
the Gaussian error term in the
plaid model is fairly robust to heavily tailed errors.

We assume that the variables $Y_{ij}$'s given the labels $(\rho,\kappa
)$ and $(\sigma^2,\Theta)$ are independent, that is,
%
%e1 #&#
\begin{equation}
P \bigl(y|\rho, \kappa,\sigma^{2},\Theta \bigr)=\prod
_{i,j}\frac{1}{\sigma}\phi \biggl(\frac{y_{ij}-\mu_{ij}(\rho,\kappa,\Theta)}{\sigma} \biggr),
\end{equation}
where $\phi$ stands for the standard normal density.
Given the bicluster labels $(\rho,\kappa)$, we define
$I_k=\{ i : \rho_{ik}=1\}$
as the set of rows in the $k$th bicluster, and
$J_k=\{j : \kappa_{jk} =1 \}$ as the set of columns in the $k$th
bicluster, $k=1,\ldots, K$.
The $k$th bicluster is given by $B_k = I_k \times J_k$.
Let $n_k$ be the number of elements in the $k$th bicluster. The number
of rows
and columns in this bicluster will be denoted by $r_{k}$ and $c_{k}$,
respectively.
Note that $n_k = r_{k} \times c_{k}$. Let $\mathbf{1}_m$ denote the
vector of all $1$'s in $\mathbb{R}^m$,
and $\mathbf{I}_{m}$ stand for the identity matrix of dimension $m$.
We further assume that, given the bicluster labels, the prior of the
gene effects $\{\alpha_{ik}\}$
is a multivariate normal distribution with mean zero
and variance--covariance matrix given by $\sigma_{\alpha}^2 V_k= \sigma
_{\alpha}^2 (\mathbf{I}_{r_k} - \frac{1}{r_k} \mathbf{1}_{r_k}
\mathbf{1}_{r_k}')$.\vspace*{2pt} As shown in \cite{Chekouo-Murua-2012},\vspace*{1pt} we may
change the
parametrization of the model to a proper multivariate normal vector
$a_k \sim N(0, \sigma_{\alpha}^2 I_{r_k})$
so that $\alpha_k = V_k a_k$.\label{def:a_k}
Similarly, we suppose that the prior for
$\{\beta_{jk}\} | (\rho,\kappa)$ follows a multivariate normal distribution
with mean zero and variance--covariance matrix given by $\sigma_{\beta
}^2 U_k= \sigma^{2}_{\beta}(\mathbf{I}_{c_k} - \frac{1}{c_k} \mathbf
{1}_{c_k}\mathbf{1}_{c_k}')$.\vspace*{2pt}
Note that these prior distributions satisfy the conditions of
identifiability in the model, that is, they ensure that
the gene and condition effects add up to zero for each bicluster.
We set zero-mean independent normal priors with variances
$\sigma^2_{\mu_0}$, and $\sigma^2_{\mu}$
for the means $\mu_0$ and $\mu$, respectively; and set
a scaled inverse chi-squared prior with scale $s^2$ and
degrees-of-freedom $\nu$ for the variance $\sigma^2$. These
hyperparameters are to be chosen
adequately. For example, in our analysis in Section~\ref
{sec:experiments}, we set $\sigma^2_{\mu_0}=\sigma^2_{\mu}=0.5$, and
$\nu=1, s^2=0.05$.

%s2.2 #&#
\subsection{A prior for the bicluster membership}
\label{sec:prior}

The gene labels $\rho_{ik}$ as well as the condition labels $\kappa
_{jk}$ are usually assumed to be independent
[\cite{Zhang2010,Gu-Liu-2008}].
More realistically, in this work, we incorporate prior knowledge on the
relation between genes and
between conditions (if applicable)
by means of relational graphs. For example, the gene relational graph
is an $r$-nearest-neighbor graph for which the nodes correspond to the
set of genes and the edges correspond to the set of ``most similar'' or
''closer'' genes.
It is this notion of similarity that contains the relational
information between genes.
We define these similarities based on the GO annotations, which define
the association between gene products and terms.
GO terms are organized in a directed acyclic graph (DAG) in which the
parent-child relationships are edges.
In this graph, a GO term can have multiple parents. All
the GO annotations associated with a term inherit all the properties of
the ancestors of those terms.
Thus, child annotations inherit annotations from multiple parent terms.
We adopt
Lin's pairwise similarity [\cite{Lin-1998}], which is based on the
minimum subsumer of \cite{Resnik-1999}, as a means to build a notion of
semantic similarity between any two
GO annotations. This idea was first introduced by \cite{Lord-et-al-2003}.
Further details can be found in the supplementary material accompanying
this paper [\cite{Chekouo-et-al-supp-2015}].
Let $d^{\rho}(i,i')=1-\operatorname{sim}(i,i')$ denote the distance between genes $i$
and $i'$ induced by Lin's similarity
between the genes $\operatorname{sim}(i,i')$.
The gene relational graph is defined as having edge weights equal to
\[
B_{ii'}\bigl(T^{\rho}, \sigma_{\rho}\bigr)=
\frac{1}{T^{\rho}}\exp \biggl(-\frac
{1}{2\sigma^{2}_{\rho}}d^{\rho}
\bigl(i,i'\bigr)^{2} \biggr). %
\]
Here, $T^{\rho}$ and $\sigma_{\rho}$ are the temperature and kernel
bandwidth parameters of the graph, respectively.
We assume that
$B_{ii'}(T^{\rho}, \sigma_{\rho})=0$ for pairs of genes not connected
by an edge in the $r$-nearest-neighbor data graph.
The larger the weights, the more similar the genes.
We will use the notation $i \sim i'$ for nodes that are connected by an
edge in the data graph.
For example, for the RD data, we fix $r=15$ to define the
$r$-nearest-neighbor graph for genes,
as this is often recommended for high-dimensional data [\cite
{Blatt1996,Stanberry-Murua-Cordes-2008}].
With a set of 4645 probe-sets of the RD data, we obtain a sparse
graph, with a total of 135{,}498 edges,
which is a total of 0.63\% connectivity in the graph.
This corresponds to an average graph degree (number of edges spawned
from each node) of 29.

The distribution of the gene labels in this graph is given by the
binary Gibbs random field
%
%e2 #&#
\begin{eqnarray}
\label{eq:h:rho} p\bigl(\rho_{k}|a,T^{\rho},
\sigma^{2}_{\rho}\bigr) &\propto& h_{\rho, k} \bigl(\rho
_k, T^{\rho}\bigr)
\nonumber
\\[-8pt]
\\[-8pt]
& \doteq& \exp \Biggl\{\sum^{p}_{i=1}a_{i}
\rho_{ik} + \sum_{i \sim
i'}B_{ii'}
\bigl(T^{\rho},\sigma_{\rho}\bigr) \mathbf{1}_{\{ \rho_{ik}=\rho_{i'k}\}
}
\Biggr\},
\nonumber
\end{eqnarray}
where $a=\{a_i\}_{i=1}^p$ are hyperparameters that control the amount
of membership ($\rho_{ik}=1$) in the bicluster,
and, for every relation $A$, $\mathbf{1}_{A}$ denotes the indicator
function that takes the value $1$ if and only if the
relation $A$ is satisfied. This Gibbs field is actually a binary
auto-logistic distribution on the labels
[\citeauthor{Besag1974} (\citeyear{Besag1974,Besag2001}), \cite{Winkler2003}].
This Gibbs prior favors biclusters formed by similar genes in the sense
of the distances or similarities
chosen to build the relational graph.

\subsubsection*{The conditions prior}

A similar prior relational graph may be built for the conditions if a
notion of similarity between the
conditions can be defined. This is the case, for example, when the
conditions correspond to similar measurements taken
over a period of time, such as in gene expression evolution (i.e.,
time-course) profiles. In this case,
the distance between conditions may incorporate a measure of smoothness
of the time-course profile
during consecutive measurements. Alternatively, a measure of
correlation may be
incorporated in the similarities if a moving average or specific ARMA
process is assumed on the time-course profiles.
These aspects of the modeling processes are better explained within the
context of specific applications, such as the ones
described in Section~\ref{sec:experiments}.
For the moment, assume that such a distance between conditions may be
defined. We denote the distance
between two conditions $j$ and $j'$ by $d^{\kappa}(j, j')$.
The condition relational graph is defined to have edge weights equal to
\[
D_{jj'}\bigl(T^{\kappa}, \sigma_{\kappa}\bigr)=
\frac{1}{T^{\kappa}}\exp \biggl(-\frac{1}{2\sigma^{2}_{\kappa}}d^{\kappa}
\bigl(j,j'\bigr)^{2} \biggr). %
\]
As before, $T^{\kappa}$ and $\sigma_{\kappa}$ are the temperature and
kernel bandwidth parameters of the graph, respectively.
And we assume that
$D_{jj'}(T^{\kappa}, \sigma_{\kappa})=0$ for pairs of conditions not
connected by an edge.
The distribution of the condition labels in this graph is then given by
the binary auto-logistic distribution
%
%e3 #&#
\begin{eqnarray}\label{eq:h:kappa}
p\bigl(\kappa_{k}|c,T^{\kappa} , \sigma^{2}_{\kappa}
\bigr) &\propto& h_{\kappa,
k} \bigl(\kappa_k, T^{\kappa}
\bigr)\nonumber\\[-8pt]\\[-8pt]
& \doteq& \exp \Biggl\{\sum^{q}_{j=1}
c_{j} \kappa_{jk} + \sum_{j \sim j'}
D_{jj'}\bigl(T^{\kappa} ,\sigma_{\kappa}\bigr)
\mathbf{1}_{\{\kappa_{jk}=\kappa
_{j'k}\}} \Biggr\}, \nonumber
\end{eqnarray}
where $c=\{c_j\}_{j=1}^q$\vspace*{2pt} are hyperparameters that control the amount
of condition membership ($\kappa_{jk}=1$) in the bicluster.
Note that in the absence of any prior information on the dependency
between conditions, we may assume
that all pairs of conditions $(j,j')$ are far apart and, consequently,
that $D_{jj'}(T^{\kappa}, \sigma_{\kappa})=0$
for all pairs $(j, j')$.
This leads to a prior where all the condition labels $\kappa_{jk}$ are
a priori independent.

%s3 #&#
\section{Posterior estimation}\label{sec:posterior}

To estimate the posterior of the parameters, especially the one
associated with the labels $(\rho,\kappa)$,
we use a hybrid stochastic algorithm.
First, an augmented model is considered in order to efficiently sample
the labels through a block Gibbs sampling. This is the Swendsen--Wang algorithm
[\cite{SwendsenWang1987}], which is well known in the physics and
imaging literature. We briefly describe it hereafter.
The effect and variance parameters are readily sampled using the usual
Gibbs sampler.
However, the temperature hyperparameters associated with the label
priors need extra consideration.
In order to sample from their posterior, one needs to know the
normalizing constant of the priors,
which are unfortunately intractable. To solve this impasse, we adopt
the Wang--Landau algorithm [\cite{WangLandau2001,AtchadeLiu2010}], which
is a technique that efficiently samples from a grid of finite
temperature values
by cleverly estimating the normalizing constant at each iteration. The
algorithm travels
efficiently over all the temperatures by penalizing each visit. The
resulting algorithm is also referred to
as a flat-histogram algorithm. Next, we further explain how the
technique is applied to our model.

%s3.1 #&#
\subsection{Sampling the labels with known temperatures}\label{sec:sampling:labels}

Let the number of biclusters $k$ be fixed.
We denote the partial residuals by
$z_{ijk}=y_{ij}-\mu_{0}-\sum_{k'\neq k}^{K}(\mu_{k'} + \alpha_{ik'} +
\beta_{jk'})\rho_{ik'}\kappa_{jk'}$.
The likelihood is given by
%
%e4 #&#
\begin{eqnarray*}
P\bigl(y|\rho,\kappa,\sigma^{2},\Theta\bigr) &\propto&\frac{1}{\sigma^{np}}
\exp \biggl\lbrace-\frac{1}{2\sigma^{2}}\sum_{i,j}
\bigl(z_{ijk}-\rho_{ik}\kappa_{jk}(
\mu_k+\alpha_{ik}+\beta_{jk})\bigr)^{2}
\biggr\rbrace
\\
& =&\frac{1}{\sigma^{np}} \exp \biggl\lbrace-\frac{1}{2\sigma^{2}}\sum
_{i,j} \rho_{ik}\kappa _{jk}(z_{ijk}
- \mu_k - \alpha_{ik} - \beta_{jk})^{2}\\
&&{}- \frac{1}{2\sigma^{2}}\sum_{i,j} (1 -
\rho_{ik}\kappa _{jk})z_{ijk}^{2} \biggr
\rbrace.
\end{eqnarray*}
Consequently, the full conditional probability of the genes' labels is
given by
\begin{eqnarray*}
P\bigl(\rho_{k}|y,\rho_{-k}, \kappa_k,
\sigma^2, \Theta, T^{\rho}\bigr)  \propto\exp \biggl\lbrace
\sum_{i}A_{ik}\rho_{ik} +\sum
_{i \sim i'}B_{ii'}\bigl(T^{\rho},
\sigma_{\rho}\bigr) \mathbf{1}_{\{\rho
_{ik} = \rho_{i'k}\}} \biggr\rbrace,
\end{eqnarray*}
where $\rho_{-k} = \rho\setminus\rho_k$ and
\[
A_{ik}= a_{i}- 0.5 \sigma^{-2}\sum
^{q}_{j=1} \bigl\{\kappa_{jk}(z_{ijk}-
\mu _k-\alpha_{ik}-\beta_{jk})^{2}-
\kappa_{jk}(z_{ijk})^{2} \bigr\}.
\]
To sample from this full conditional, we use the
Swendsen--Wang algorithm [\cite{SwendsenWang1987}].
This algorithm samples the labels in blocks by taking into account the
neighborhood system of the data graph.
It defines a set of the independent auxiliary $0$--$1$ binary variables
$R=\{R_{ii'} : i,i'=1, \ldots, p\}$,
called the bonds. The bonds are set to $1$ with label-dependent
probabilities given by
%
%e5 #&#
\begin{equation}
p_{ii'} \doteq P(R_{ii'}=1|\rho_{k})= \bigl(1-\exp
\bigl\{-B_{ii'}\bigl(T^{\rho}, \sigma_{\rho}\bigr)\bigr\}
\bigr) \mathbf{1}_{\{\rho_{ik} = \rho_{i'k}\}} \mathbf{1}_{ \{i \sim i'\}}.
\end{equation}
The bond $R_{ii'}$ is said to be \textit{frozen} if $R_{ii'}=1$.
Note that necessarily a frozen bond can occur only between neighboring
points that share the same label.
A set of data graph nodes is said to be connected if, for every pair of
nodes $(i, i')$ in the set,
there is a path of frozen nodes in the set connecting $i$ with $i'$.
The Swendsen--Wang algorithm is used to sample the labels as follows:
\begin{longlist}[2.]
\item[1.] Given the labels $\rho_{k}$, each bond $R_{ii'}$ is frozen
independently of the others with probability
$p_{ii'}$ if $i\sim i'$ and $\rho_{ik}=\rho_{i'k}$. Otherwise, the bond
is set to zero.

\item[2.] Given the bond variables $R$, the graph is partitioned into its
connected components.
Each connected component $C$ is randomly assigned a label. The
assignment is done independently,
with $1$-to-$0$ log-odds equal to $\sum_{i\in C}A_{ik}$. In the
special case of the Ising model
and, more generally, when $A_{ik}=0$ for all $i$, the labels are chosen
uniformly at random.
\end{longlist}
Given the gene labels, the condition labels are sampled in a similar way.

%s3.2 #&#
\subsection{Sampling the labels with unknown temperatures}\label{sec:sampling:temperatures}

We assume that the temperatures $T^{\rho}$ and $T^{\kappa}$ take a
finite number of values.
Let ${\cal T}_{\rho}$ and ${\cal T}_{\kappa}$ be the sets of $m$ and
$n$ possible values
for $T^{\rho}$ and $T^{\kappa}$, respectively.
We assume that the prior distribution of $(T^{\rho}, T^{\kappa})$ is a
uniform distribution on the grid of values
${\cal T}_{\rho} \times{\cal T}_{\kappa}$.
Note that $p(\sigma^2, \Theta, \rho, \kappa, T^{\rho}, T^{\kappa}|y )$
is directly proportional to
\[
p\bigl( y | \sigma^2, \Theta, \rho, \kappa\bigr) \pi\bigl(
\sigma^2, \Theta\bigr) \prod_{k=1}^K
\biggl( \frac{h_{\rho, k}( \rho_k, T^{\rho} )}{Z_{\rho
}(T^{\rho})} \frac{h_{\kappa, k}( \kappa_k, T^{\kappa} )}{Z_{\kappa
}(T^{\kappa})} \biggr), %
\]
where
$Z_{\rho}(T)$ and $Z_{\kappa}(T)$ denote the normalizing constants for
$h_{\rho, k}( \rho_k, T )$ and
$h_{\kappa, k}( \kappa_k, T^{\rho} )$, respectively [see equations (\ref
{eq:h:rho}) and (\ref{eq:h:kappa})].
In general, these constants cannot be easily evaluated
and are intractable, except for the very simplest cases.
MCMC techniques, such as Metropolis--Hastings,
are of no use here because the constants change with the value of $T$.
Instead, in order to obtain samples from the posterior of the labels,
we use a stochastic algorithm based on the Wang--Landau algorithm
[\cite{WangLandau2001,AtchadeLiu2010}].
The sampling from this algorithm simultaneously provides approximate
samples from the posterior
of the labels and the parameters $(\sigma^2, \Theta)$ and
estimates of the posterior probability mass function of $(T^{\rho},
T^{\kappa})$.
\cite{AtchadeLiu2010} provided a nice
exposition of the algorithm and showed its convergence. \cite{Murua-Wicker-2012}
successfully used a variant of the Wang--Landau algorithm to estimate
the posterior of the temperature
of the Potts model.
The Wang--Landau algorithm considers the target joint distribution
%
%e6 #&#
%e7 #&#
\begin{eqnarray}\label{eq:joint:target}
\quad&&\pi\bigl( \sigma^2, \Theta, \rho, \kappa, T^{\rho},
T^{\kappa} \bigr) \nonumber\\[-8pt]\\[-8pt]
\quad&&\qquad\propto
p\bigl( y | \sigma^2, \Theta, \rho, \kappa\bigr) \pi\bigl(
\sigma^2, \Theta\bigr) \prod_{k=1}^K
h_{\rho, k}\bigl( \rho_k, T^{\rho} \bigr)
h_{\kappa, k}\bigl( \kappa_k, T^{\kappa} \bigr) / \psi\bigl(
T^{\rho}, T^{\kappa}\bigr), \nonumber
\end{eqnarray}
where $\psi( T^{\rho}, T^{\kappa})$ is given by
%
%e8 #&#
\begin{equation}
\label{eq:joint:target:normalizing}
\hspace*{15pt}
Z^{-1} \sum_{\rho, \kappa} \int p\bigl( y |
\sigma^2, \Theta, \rho, \kappa\bigr) \pi\bigl( \sigma^2,
\Theta\bigr)\,d\sigma^2\,d\Theta \prod_{k=1}^K
h_{\rho, k}\bigl( \rho_k, T^{\rho} \bigr)
h_{\kappa, k}\bigl( \kappa_k, T^{\kappa} \bigr),\hspace*{-5pt}
\end{equation}
where $Z$ is the constant such that
$\sum_{T^{\rho} \in{\cal T}_{\rho}, T^{\kappa} \in{\cal T}_{\kappa} }
\psi( T^{\rho}, T^{\kappa}) = 1$.
The algorithm samples from iterative stochastic approximations
of this distribution (see the algorithm steps below),
so that the marginal of the parameters and labels
converges to the target marginal
$\pi(\sigma^2, \Theta, \rho, \kappa) = p( \sigma^2, \Theta, \rho,
\kappa| y)$
and
the marginal of $( T^{\rho}, T^{\kappa})$ converges to
$\pi(T^{\rho}, T^{\kappa})$, which turns out to be
a uniform distribution on the grid of temperatures ${\cal T}_{\rho}
\times{\cal T}_{\kappa}$.
The main idea of the stochastic approximation is to replace $\psi(
T^{\rho}, T^{\kappa})$
by an iterative estimate, say $\hat{\psi}( T^{\rho}, T^{\kappa})$.
Consider equation (\ref{eq:joint:target}) with
$\psi( T^{\rho}, T^{\kappa})$ replaced by its estimate $\hat{\psi}(
T^{\rho}, T^{\kappa})$.
Since $\pi(T^{\rho}, T^{\kappa})$ is uniform, then integrating this equation
so as to obtain the estimate $\hat{\pi}(T^{\rho}, T^{\kappa})$, and
using equation (\ref{eq:joint:target:normalizing}),
we have that
at convergence
%
%e9 #&#
\begin{eqnarray}
\frac{\hat{\psi}( T^{\rho}, T^{\kappa})}{\sum_{t^{\rho} \in{\cal
T}_{\rho}, t^{\kappa} \in{\cal T}_{\kappa} }
\hat{\psi}( t^{\rho}, t^{\kappa})}  \approx\psi\bigl( T^{\rho}, T^{\kappa}\bigr).
\label{eq:phi}
\end{eqnarray}
Therefore, the quantities given in the left-hand side of equation (\ref
{eq:phi}) give
an estimate of the posterior probability mass function of the
temperatures $( T^{\rho}, T^{\kappa})$.

Let ${\cal T}_{\rho} = \{t_1 < t_2 < \cdots< t_m\}$ be the set of
temperatures considered.
The Wang--Landau algorithm we have implemented depends on an updating
proposal of the form
$q( T^{\rho}, T^{\kappa} | T^{\rho, (t)}, T^{\kappa, (t)} ) = q_{\rho}(
T^{\rho} | T^{\rho, (t)} )
q_{\kappa}( T^{\kappa} | T^{\kappa, (t)} )$,
with
$q_{\rho}( t_1, t_2 ) = q_{\rho}( t_m, t_{m-1}) = 1$
and $q_{\rho}(t_i,t_{i-1})=q_{\rho}(t_{i},t_{i+1})=0.5$ if $1<i<m$.
The proposal $q_{\kappa}$ is similarly defined.
This proposal corresponds to the proposal of \cite{GeyerThompson1995}
that was used within the context of simulated tempering.
\cite{AtchadeLiu2010} suggested a different proposal based on a
multinomial distribution.
However, their proposal involves considerably more
computation.

The algorithm proceeds as follows:
Given $(\sigma^{2, (t)},\Theta^{(t)}, \rho^{(t)}, \kappa^{(t)}, T^{\rho
, (t)},T^{\kappa, (t)})$ and
$\hat{\psi}^{(t)}= \{ \hat{\psi}( t^{\rho}, t^{\kappa}) : ( t^{\rho},
t^{\kappa}) \in{\cal T}_{\rho} \times{\cal T}_{\kappa} \}$
at iteration $t$:
\begin{longlist}[(iii)]
\item Sample $T$ from the proposal distribution $q_{\rho}(\cdot|
T^{\rho, (t)})$.
Set $T^{\rho, (t+1)}=T$ with probability
\[
\min \Biggl( 1, R_{\rho}(T) \exp \Biggl\{ \sum
_{k=1}^{K}\sum_{i \sim i'}
\bigl(B_{ii'}\bigl(T,\sigma^{2}_{\rho
}
\bigr)-B_{ii'}\bigl(T^{\rho, (t)},\sigma^{2}_{\rho}
\bigr)\bigr) \mathbf{1}_{\{ \rho_{ik}^{(t)}=\rho_{i'k}^{(t)} \}} \Biggr\} \Biggr), %
\]
otherwise set $T^{\rho, (t+1)}=T^{\rho, (t)}$,
where $R_{\rho}(T) = \frac{q_{\rho}(T|T^{\rho, (t)})}{q_{\rho}(T^{\rho,(t)}|T)}
\frac{\hat{\psi}( T^{\rho, (t)}, T^{\kappa, (t)})}{\hat{\psi}( T,
T^{\kappa, (t)})}$.
\item Sample $T$ from the proposal distribution $q_{\kappa}(\cdot|
T^{\kappa, (t)})$.
Set $T^{\kappa, (t+1)}=T$ with probability
\[
\min \Biggl( 1, R_{\kappa}(T) \exp \Biggl\{ \sum
_{k=1}^{K}\sum_{j \sim j'}
\bigl(D_{jj'}\bigl(T,\sigma^{2}_{\kappa
}
\bigr)-D_{jj'}\bigl(T^{\kappa, (t)},\sigma^{2}_{\kappa}
\bigr)\bigr) \mathbf{1}_{\{ \kappa_{jk}^{(t)}=\kappa_{j'k}^{(t)}\} } \Biggr\} \Biggr), %
\]
otherwise set $T^{\kappa, (t+1)}=T^{\kappa, (t)}$, where $R_{\kappa}(T)
=\frac{q_{\kappa}(T|T^{\kappa, (t)})}{q_{\kappa}(T^{\kappa,(t)}|T)}
\frac{\hat{\psi}( T^{\rho, (t)}, T^{\kappa, (t)})}{\hat{\psi}( T^{\rho,
(t)}, T)}$.

\item\label{psi1} Update $\hat{\psi}^{(t+1)}$: for $( t^{\rho},
t^{\kappa}) \in{\cal T}_{\rho} \times{\cal T}_{\kappa}$, set
%
%e10 #&#
\begin{eqnarray}\label{eq:log:psi}
&&\log\hat{\psi}^{(t+1)}\bigl(t^{\rho}, t^{\kappa}\bigr)\nonumber\\[-8pt]\\[-8pt]
&&\qquad=\log
\hat{\psi }^{(t)}\bigl(t^{\rho}, t^{\kappa}\bigr) +
\gamma^{(t)} \biggl( \mathbf{1}_{ \{ (T^{\rho,(t+1)}, T^{\kappa,
(t+1)}) =(t^{\rho}, t^{\kappa})\} }-\frac{1}{m n}
\biggr). \nonumber
\end{eqnarray}
\item Sample $\rho^{(t+1)}$ and $\kappa^{(t+1)}$ with the Swendsen--Wang
algorithm.
\item\label{par} Sample $(\sigma^{2, (t+1)}, \Theta^{(t+1)})$ using
the usual Gibbs sampler.
\end{longlist}
In step (\ref{psi1}), $\gamma^{(t)}$ is a random sequence of real
numbers decreasing slowly to 0.
We chose $\gamma^{(t)}$ according to the Wang--Landau schedule suggested
by \cite{AtchadeLiu2010}.
The sequence $\gamma^{(t)}$ is kept constant until the histogram of the
temperatures is flat, that is, until $(T^{\rho, (t)}, T^{\kappa, (t)})$
has equiprobably visited all the values of the grid ${\cal T}_{\rho}
\times{\cal T}_{\kappa}$.
At the $k$th recurrent time $n_k$ such that $(T^{\rho, (t)}, T^{\kappa, (t)})$
is approximately uniformly distributed, we set $\gamma
^{(n_{k}+1)}=\gamma^{(0)}/2^k$ where $\gamma^{(0)}=1$.
When $\gamma^{(t)}$ becomes too small, $\gamma^{(t)}$ is set to
$0.0001/t^{0.7}$.
In practice, a very large number of iterations is needed to reach
convergence of the quantities
given in equation~(\ref{eq:phi}) [or equation~(\ref{eq:log:psi})].
We carried out a small simulation (not shown here) to get a better idea
of the number of simulations needed for a problem
like ours. The answer lies at about one-half million iterations.
A theoretical proof of the convergence of this algorithm is given in
the supplementary material [\cite{Chekouo-et-al-supp-2015}].

In step (\ref{par}), the parameters $(\sigma^2, \Theta)$ are sampled
with a Gibbs sampler.
The full conditional posterior of the parameters $(\sigma^2,\Theta)$ is
straightforward to
derive; hence, it is not spelled out here.
The temperatures ${\cal T}_{\rho}$ (and also the set ${\cal T}_{\kappa
}$, if appropriate)
are obtained by using the procedure of \cite{Murua-Wicker-2012} to
elicit their prior critical
temperatures from the random cluster models associated with the Potts model.
The kernel bandwidth parameters $\sigma_{\rho}$ and $\sigma_{\kappa}$
are kept constant and set to the corresponding average nearest-neighbor distance
[\cite{Blatt1996}].

%s4 #&#
\section{Experiments with simulated data}\label{sec:experiments}

To build our simulated data sets, we used
two different pools of genes: one from
the yeast cycle data [\cite
{Cho-et-al-1998,Lichtenberg-et-al-2005,Rustici-et-al-2004}]
and the second from the retinal detachment (RD) data [\cite{GEO-2002}].

The yeast cycle data set shows the time-course fluctuation of the
log-gene-expression-levels of 6000 genes over 17 time points.
The data have been analyzed by several researchers
[\citeauthor{Cho-et-al-1998} (\citeyear{Cho-et-al-1998,Cho-et-al-2004}), \citet{Mewes1999,Tavazoie1999}]
and are a classical example for testing clustering algorithms [\cite
{Yeung-et-al}].
We use the five-phase subset of this data, which consists of $384$ genes
with expression levels that peak at different time points, corresponding
to the five phases of the cell cycle.
Of the $384$ genes, only $355$ are annotated with GO terms.

The RD data set is described in greater detail in Section~\ref
{sec:applications:RD}.
We used this data set so as to have simulations that resemble the RD
data more closely.
We randomly chose 2000 probe-sets (i.e., genes) out of the 4645
probe-sets present in these data
in order to study many scenarios for the simulated data.

Based on Lin's pairwise similarities, discussed in Section~\ref
{sec:prior}, we built
corresponding relational graphs comprising the annotated genes.
As with the real data, we simulated 38 conditions for the genes taken
from the RD data set.
Recall that the RD data set consists of a group of 19 biopsies from
patients with RD and a control group of 19 non-RD biopsies.
As described in \cite{Delyfer-2011}, the patients can be further
organized into three classes of RD:
early stage (RD $\leq1$ month, 5 patients), mid-term stage ($1$ month
$>$ RD $\leq3$ months, 7 patients) and late stage (RD $> 3$ months, 7
patients).
The relational condition graph associated with the genes from the RD
data set was built so that
patients in the same group were related in the graph.
The distances between patients in the same group were assumed to be
the same.

For the genes taken from the yeast cycle data set,
we simulated 17 conditions, the same number of conditions found in the
real data.
The modeling of the relational condition graph associated with these genes
was inspired by the time dependency in the data.
This allowed us to consider biclusters formed by consecutive
conditions, which are
easier to visualize. Thus, for these simulated data, the similarity
between conditions
was induced by the correlation $\xi$ between time-consecutive conditions.
The correlation distance between conditions was set to
\[
d^{\kappa}\bigl( j, j'\bigr) = %
\cases{ 2 \bigl( 1
- \xi^{|j- j'|}\bigr),\quad & $\bigl|j- j'\bigr| \leq3$,\vspace*{2pt}
\cr
0 & otherwise.}
\]
The value of the correlation parameter does not affect the relational
structure given by the $r$-nearest-neighbor graph. In our simulations,
we set $\xi= 0.8$.
Setting $\xi$ as an unknown parameter of the model would unnecessarily
complicate the
model because conducting inference on $\xi$ would involve knowledge of
the normalizing
constant, which in turns depends on $\xi$ and the temperature.
A high value of $\xi$ should guide the model to consider clustering
time-consecutive
conditions together.

As our label prior favors common labels for genes or conditions that
are strongly related
in the graph, we used a hierarchical clustering (e.g., Ward's minimum
variance method [\cite{Ward-1963}])
with different tree cutoffs to generate labels for
different numbers of biclusters. Clusters that split at higher cutoffs
in the tree were used as candidates
for overlapping biclusters.

\begin{figure}

\includegraphics{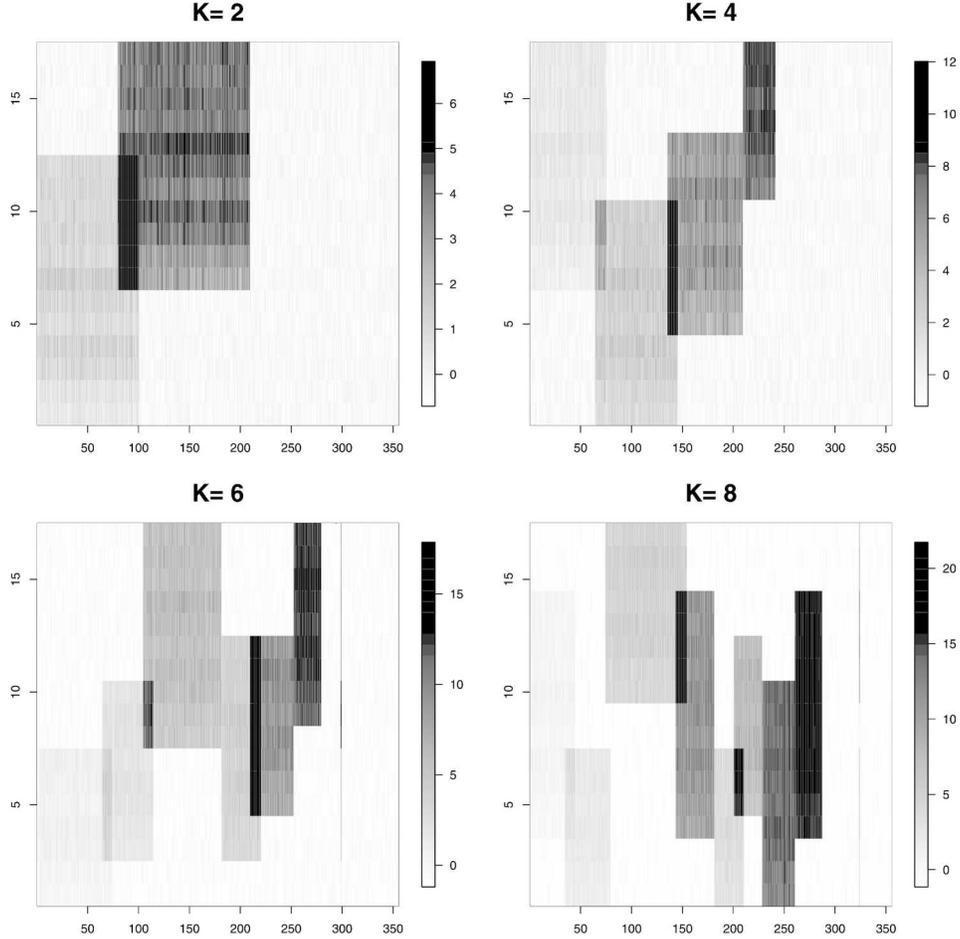}

\caption{Examples of simulated data.}
\label{fig:simulated:data}
\end{figure}

The expression levels of the bicluster cells associated with the data
for genes taken from the
yeast cycle data were generated as follows:
$\mu_{0}$ was generated from a Normal$(0, 0.05)$ distribution;
$\mu_{k}$ was generated from a Normal$(2(k+1),0.05)$, $k=1,2,\ldots,K$
distribution;
the gene effects $\alpha_{ik}$ were generated
as normal
distributions, with the means equal to $\mu_{\alpha_{ik}}=\frac
{2}{1+\exp(-i)}-
\frac{1}{r_{k}}\sum_{i}\frac{2}{1+\exp(-i)}$,
and the variances equal to their prior variances,
while keeping the constraint $\sum_{i=1}^{p}\alpha_{ik}\rho_{ik}=0$,
$k=1,\ldots, K$
(see the last paragraph of Section~\ref{sec:plaid:model} on
page~\pageref{def:a_k});
the condition effects $\beta_{jk}$ were generated similarly; and the variance
$\sigma^{2}$ was generated from an inverse-$\chi^{2}(3,0.03)$.
In this fashion, we created data sets with the following numbers of
biclusters: $K=2,3,4,5,6,7, 8$.
Each of these cases was replicated 15 times.
Figure~\ref{fig:simulated:data} shows some examples of the simulated
data for different values of~$K$.

The expression levels of the bicluster cells associated with the data
for genes taken from the RD data set
were generated in the same manner, except for the parameters $\mu_k$
that were generated from a Normal distribution, with mean
$2(10(k+1)/K + 1)$ and variance $0.05$. In this case, we created data
sets with the following numbers of
biclusters: $K=4, 8, 16, 24, 30, 40, 50$.
Each of these cases was replicated 15 times.

%f1 #&#

%s4.1 #&#
\subsection{The F1-measure of performance}

A measure of similarity between two sets of biclusters
$M_1=\{ A_1,\ldots, A_k\}$ and
$M_{2}=\{B_1, \ldots, B_{\ell} \}$ is given
by the so-called F1-measure [\cite{Santamaria-et-al-2007,Turner-et-al-2005}].
The F1-measure is an average between {\em recall} and {\em precision},
two measures of
retrieval quality introduced in the text-mining literature [\cite
{Allan-et-al-1998}].
Let $A, B$ be two biclusters, $r_A$ and $r_{B}$ be the number
of genes in $A$ and $B$, $c_A$ and $c_{B}$ be the number of conditions
in $A$ and $B$,
and $n_A= r_A c_A$ and $n_{B}=r_{B}c_{B}$
be the number of elements in $A$ and $B$, respectively. Precision and
recall are given by
\[
\mathrm{recall} =\frac{(r_{A\cap B})(c_{A\cap B})}{n_{B}}, \qquad \mathrm{precision} =\frac{(r_{A\cap B})(c_{A\cap B})}{n_{A}}.
\]
Recall is the proportion of elements in $B$ that are in $A$.
Precision is the proportion of elements in $A$ that are also found in $B$.
The F1-measure between $A$ and $B$ is given by
$F_1(A,B)=2 (r_{A\cap B}) \times(c_{A\cap B}) / (n_{A}+n_{B})$.
When several target biclusters (or estimated biclusters) $M_1$ are to
be compared
with known biclusters $M_{2}$,\vspace*{2pt} we use the F1-measure average:
$F_1(M_1,M_2)=\frac{1}{k}\sum_{i=1}^k\max_{j}F_1(A_i,B_j)$. The
estimated biclusters $M_1$ are obtained by using a threshold of 0.5 on
the marginal posterior probabilities of the labels from our stochastic
algorithm.

%s4.2 #&#
\subsection{Comparison results}
We show the results of a performance comparison between the Gibbs-plaid
model and the Bayesian penalized plaid model of
\cite{Chekouo-Murua-2012} for each number of biclusters considered.
The penalized plaid model uses a parameter $\lambda$, which controls
the amount of overlap of the biclusters. It extends the original plaid
model of \cite{Lazzeroni-Owen-2002} and the nonoverlapping model of
\cite{Cheng-Church-2000}, which arise as special cases of the penalized
model when $\lambda$ is set to zero and infinity, respectively.
The case of $\lambda=0$ is also equivalent to our Gibbs-plaid model
when the temperatures tend toward infinity
(i.e., a model without prior interaction between the genes or between
the conditions).
\cite{Chekouo-Murua-2012} fit their model with
a Gibbs sampler, and showed that its performance
is much better than the performance of five other competitive
biclustering methods: the SAMBA algorithm of \cite{Tanay2002},
the improved plaid model of \cite{Turner-et-al-2005}, the algorithm of
\cite{Cheng-Church-2000},
the spectral algorithm of \cite{Kluger2003}, and the FABIA procedure of
\cite{Hochreiter2010FABIA}.
In this section, we extend this performance comparison by (a) including
our Gibbs-plaid model,
(b) considering a larger and much more diverse pool of genes in the
generation of data sets, and by
(c) considering a larger number of biclusters in the simulations.

%f2 #&#
\begin{figure}

\includegraphics{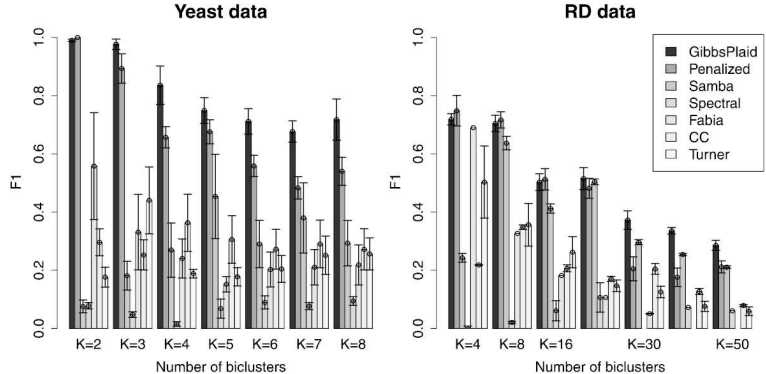}

\caption{F1-measure means.
The darkest bars correspond to our Gibbs-plaid model and the other bars
to other biclustering algorithms.
The segments on top of the bars represent plus or minus two standard deviations
estimated from 15 replicates. CC stands for the Cheng and Church algorithm.}
\label{fig:yeast:RD:simulation}
\end{figure}

The Gibbs-plaid model was run with the stopping criterion suggested by
\cite{AtchadeLiu2010}, but with the maximum number of iterations
fixed at $500{,}000$.
The penalized plaid model was run for $20{,}000$ iterations.
For both models, we used the last $10{,}000$ samples after the burn-in
period to perform the analysis and comparisons.
We set the hyperparameters of the variables $\Theta$ and $\sigma^2$ as follows:
$\sigma^2_{\mu_0}=\sigma^2_{\mu}=\sigma^2_{\alpha}=\sigma^2_{\beta}=0.5$,
$\nu=1$ and $s^2=0.05$.
Figure~\ref{fig:yeast:RD:simulation} shows the results.
Overall, the Gibbs-plaid model performed better than the penalized
plaid model and the other five biclustering algorithms. The difference
in performance was much larger when the number of biclusters was large
($K\geq30$ for the RD data and $K\geq6$ for the yeast data).
We stress that these results apply to a large simulation involving very
different pools of genes and types of conditions.
Note that with the RD data, the FABIA algorithm did not work for cases
with a large number of biclusters ($K\geq40$),
and that the spectral algorithm did not find any biclusters for all
cases (data set replicates) with $K=4$ and $K=30$.
Moreover, for $K=4, 8$ and $30$,
FABIA found biclusters in only a single case out of 15 replicates.
Similarly, for $K=40$ and $50$, the spectral algorithm found biclusters
in only a single case.

%s4.3 #&#
\subsection{Choosing the number of biclusters}\label{sec:DIC}

\begin{figure}

\includegraphics{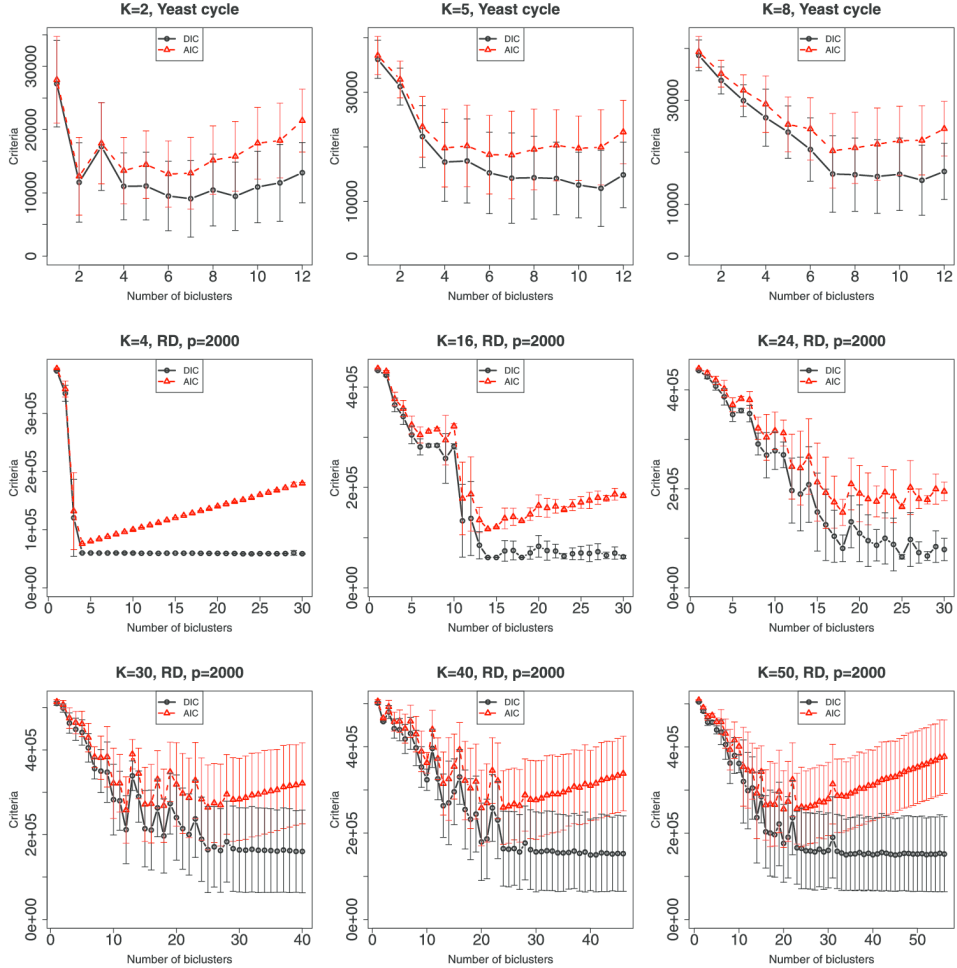}

\caption{Simulated data. Average AIC and DIC$_c$ for the Gibbs-plaid model.
The top row shows the results associated with the data sets generated
for genes from the yeast cycle data ($p=355$, $q=17$).
The middle and bottom rows show the results associated with the data
sets generated for genes from the RD data ($p=2000$, $q=38$).
The bars correspond to plus or minus two standard deviations.}
\label{fig:DIC:Yeast:RD}
\end{figure}

As in the work of \cite{Chekouo-Murua-2012}, we used
two model selection criteria to decide on the appropriate number of
biclusters for each data set.
We used the AIC [\cite{Akaike-1974}] and the
conditional DIC (DIC$_c$), which was considered in \cite{Chekouo-Murua-2012}
and is given by
\begin{eqnarray*}
\mathrm{DIC}_{c} &=&-2E_{\sigma^2,\Theta, \rho, \kappa} \bigl[\log p\bigl(y|\sigma
^2, \Theta, \rho, \kappa\bigr)|y \bigr]+ p_{c}\bigl(
\tilde{\sigma}^2, \tilde{\Theta}, \tilde{\rho}, \tilde{\kappa} \bigr),
\end{eqnarray*}
where $(\tilde{\sigma}^2, \tilde{\Theta}, \tilde{\rho}, \tilde{\kappa
})$ is the maximum a posteriori
estimator of $(\sigma^2, \Theta, \rho, \kappa)$ and
\begin{eqnarray*}
&&p_c\bigl(\tilde{\sigma}^2, \tilde{\Theta}, \tilde{
\rho}, \tilde{\kappa}\bigr)\nonumber\\[-8pt]\\[-8pt]
&&\qquad = -2E_{\sigma^2,\Theta, \rho, \kappa} \bigl[\log p\bigl(y|
\sigma^2, \Theta,\rho,\kappa\bigr)|y \bigr]+2\log p\bigl(y|\tilde {
\sigma}^2, \tilde{\Theta}, \tilde{\rho}, \tilde{\kappa}\bigr),\nonumber
\end{eqnarray*}
is the corresponding effective dimension.
We computed the DIC$_c$ and AIC criteria
for all the simulated data for different values of $K$.
For the data generated from the yeast cycle data, we computed these criteria
for $k\leq12$ biclusters. For the data generated with the RD data, we
computed these criteria
for $k\leq30$ biclusters when $K\leq24$, for $k\leq36$ when $K=30$,
for $k\leq46$ when $K=40$, and for $k\leq56$ when $K=50$.
Figure~\ref{fig:DIC:Yeast:RD} shows the model selection results for
some of the simulated data sets.
We note that, in general, AIC and DIC$_c$ chose the same models for the
small data sets generated with the
pool of genes of the yeast cycle data. However, for the larger data
sets generated with the pool of genes
of the RD data, AIC tended to reach a minimum before DIC$_c$ did,
largely underestimating the true number of biclusters.
This suggests an over-penalization of complex models by AIC due to the
large number of parameters
induced by the large number of genes in the data sets. This behavior of
AIC has been noticed before
[\cite{Chekouo-Murua-2012}].
On the other hand, the elbow of the DIC$_c$'s curve (that is, the start
of the flattening of the DIC$_c$'s trajectories)
tended to occur at or after the minimum of the corresponding AIC curves.
In some cases, the DIC$_c$ criterion reached a minimum at a number of
biclusters that was larger than the true
number of biclusters.
A closer look at the extra biclusters revealed that they were, in general,
very small, containing only a couple of conditions or a handful of genes.
In addition, at the flattening of the DIC$_c$ curve, the DIC$_c$'s
values were not (statistically)
significantly different when we considered the errors in the DIC$_c$'s
estimates (the vertical segments
crossing the curve correspond to plus or minus two standard deviations;
the standard deviations
were estimated from 15 replicates).
Therefore, a possible rule of thumb is to select the biclustering model
associated
with a point in the flat part of the DIC$_c$ curve that falls near
the elbow of the
curve.
This is the rule we applied in the simulations and in the application
to a real data set, described hereafter.

%f3 #&#

%s5 #&#
\section{Application to the retinal detachment disorder data}\label{sec:applications:RD}

\begin{figure}[b]

\includegraphics{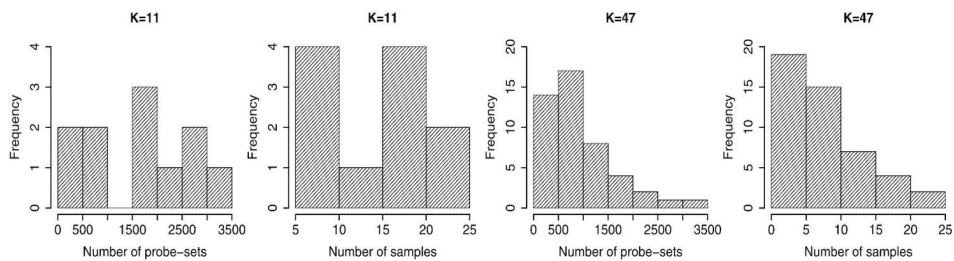}

\caption{The retinal detachment data.
The two leftmost histograms show the number of genes per bicluster (far left)
and the number of experimental conditions per bicluster (second from
the left) associated with
the solution preferred by AIC.
The rightmost histograms show the same type of information
associated with the solution preferred by DIC.}
\label{fig:RD:hist}
\end{figure}

In this section we show the application of our biclustering approach to
the data gathered from a study in which 19 biopsy samples of RD were
compared to 19 normal retinal samples
[\cite{Delyfer-2011}]. The data are available at NCBI/GEO as GSE28133
[\cite{GEO-2002}].
The first step in microarray analysis consists in filtering for
potentially relevant alterations in expression levels
and removing any changes presumably due to the inherent noise of the system
[\cite{Calza-et-al-2007,Hackstadt-Hess-2009,Gentleman-et-al-2005}].
Such filtering aims at eliminating all genes whose expression
measurements are very low,
and to whom the resulting measures can be associated with random noise
at detection-limit.
In our case, \cite{Delyfer-2011} points out that
the data is well described as a bimodal distribution where the first
peak is associated with
nonexpressed genes (i.e., where random noise at detection-limit was captured).
In order to separate the random noise peak from the second peak of
the bimodal distribution,
we followed the exact same preprocessing procedure of \cite
{Delyfer-2011} and
applied a threshold of 31.5 expression units to the expression data.
Only 32\% of all probe-set expression values in the data were retained
after the application of the threshold.
Fundamentally, this filtering step follows the belief that a gene
which is not expressed in any of the samples studied
cannot present changes in expression rates in some samples and,
therefore, all changes in the measures
are due to random noise. Therefore, we filtered out the
genes/probe-sets with very low or constant expression values along all
samples, which allowed us to concentrate on the highly reliable changes
in the transcriptome, reduce the overall
noise, and accelerate the subsequent calculations.
A further gene filtering step was done based on the intuitive belief
that if a gene expression standard deviation is too small,
then the gene may have little discriminating strength (e.g., to
discriminate between RD patients from healthy control ones)
and will be less likely to be selected.
We studied the effects of performing this preprocessing step in a
simulation study (not shown here).
We noticed that noisy genes not only increased the computational
burden, but could also decrease the biclustering performance.
After this filtering step, we obtained a data set of 4645 probe-sets
with information for 3182 different genes
(multiple probe-sets may correspond to a single gene).
We fit the Gibbs-plaid biclustering model to these data.
The DIC criterion chose 47 biclusters, a value close to the elbow,
whereas the AIC criterion chose 11 biclusters, the value of the minimum AIC.
The size of the biclusters are shown in a series of histograms in
Figure~\ref{fig:RD:hist}.

%f4 #&#

The DIC biclustering yielded a total of 20 biclusters that contained
more than 80\% of the RD samples,
and 6 biclusters that contained more than 80\% of the non-RD samples.
In contrast, the AIC biclustering yielded only 5 biclusters that
contained more than 80\% of the RD samples,
and 3 biclusters that contained more than 80\% of the non-RD samples.
Of the 20 DIC-yielded biclusters with at least 80\% of the RD samples,
18 contained 90\% of the RD samples, and 15 contained only RD samples
(i.e., they were purely RD sample biclusters).
We are particularly interested in the ``{\em significant}'' biclusters
because genes involved in these biclusters can
be viewed as biomarkers that discriminate between the patients with RD
and those without RD. In what follows, we refer to
the biclusters that contain at least 80\% of the RD samples or at least
80\% of the non-RD samples as {\em significant biclusters.}
Of particular interest are DIC biclusters 4, 41 and 6, which
respectively consist of 95\%, 91\% and 84\% of the RD samples.

\begin{figure}[b]

\includegraphics{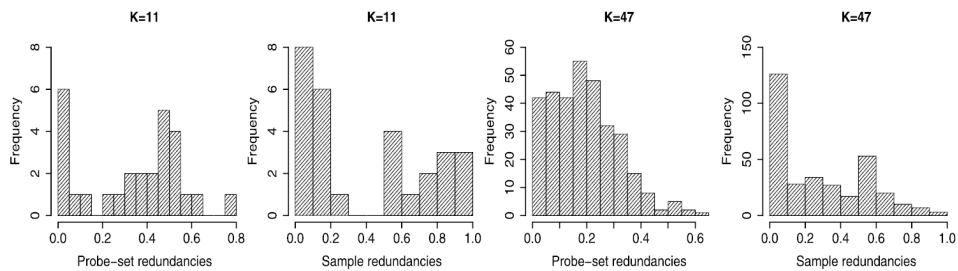}

\caption{Relative redundancy from the retinal detachment data.
The two leftmost histograms show the relative redundancy of genes
between biclusters (far left)
and the relative redundancy of samples between biclusters (second from
the left) associated with
the solution preferred by AIC.
The two rightmost histograms show the same type of information
associated with the solution preferred by DIC.
Only significant biclusters were involved in the calculations.}
\label{fig:RD:Hist}
\end{figure}

The degree of biclustering overlap and association among the
significant biclusters
may be better studied by computing the
amount of shared elements (either probe-sets or samples)
between each pair of biclusters. We computed the relative {\em
redundancy} between each pair of biclusters
as the average of the two ratios given by the
number of shared elements and the corresponding bicluster sizes.
As the DIC produced a larger
number of smaller biclusters, the corresponding results of
biclustering showed less
overlap (i.e., lower relative redundancy) than the AIC results (see
Figure~\ref{fig:RD:Hist}).
%
%f5 #&#

A more detailed inspection of the biclustering results
(see the supplementary material [\cite{Chekouo-et-al-supp-2015}] for
complete biclustering results)
revealed that those produced using DIC contained the most interesting
enrichment of GO ontologies related to photoreceptor cells (i.e., GO
ontologies ``GO:0009416 response to light stimulus'' or further
specialized branches of the previous GO term, such as ``GO:007603
phototransduction, visible light''), which were found in DIC bicluster
4 and somehow weaker in bicluster 6
(DIC biclusters 4 and 6 have a relative gene redundancy of 51.8\%).
Some other interesting biclusters showed either enrichment of GO
ontology terms for inflammatory response
(bicluster 41, which consists of 91\% RD samples) or for cell death
(bicluster 8, which consists of only 54\% RD samples).
Both types of responses have been previously described [\cite
{Delyfer-2011}], but are not related to photoreceptor cells and are
therefore less helpful in establishing a better understanding of the
fate of photoreceptor cells.
The biclusters obtained using AIC had globally similar results with
respect to enriched GO ontologies.
However, the terms related to vision and photoreceptor cells showed
less dominant enrichment.
In addition, this biclustering contains only a few ``significant'' biclusters.
Moreover, following our simulation results, the large difference in the
number of biclusters suggested by AIC and DIC indicate that the DIC
results should be more reliable than those obtained from AIC in this
case. Therefore, in the subsequent analysis, we focused on the results
obtained using DIC and, in particular,
on bicluster 4, which contained all the RD samples and only one non-RD sample.

Subsequent inspection of the protein interaction map\footnote{%
In analogy to maps of urban public transport (in particular, subway maps),
networks of protein-protein interaction have been called ``interaction maps.''
Both types of graphs have nodes that are interconnected [proteins are
connected with other proteins
when they have previously been identified to interact
(biologically/physically) with each other], and,
in both types of graphs, some nodes have a high number of connections
while the majority has simply one or two connections.}
for the proteins identified in DIC
bicluster 4 (formed by 332 probe-sets and representing 301 different proteins)
was performed using the STRING database of documented protein-protein
interactions
[\cite{Jensen-et-al-2009}].
This is displayed in Figure~\ref{fig:Bicluster4:network} (see the
supplementary material [\cite{Chekouo-et-al-supp-2015}] for a
high-resolution image).
On the basis of 301 proteins, we obtained a fairly small network of 50
directly interconnected proteins.
We decided to construct an extended network by adding proteins that
allowed us to link two or more of the 301 proteins
from bicluster 4, and for which the expression values were sufficiently
high to call them unambiguously expressed genes.
Again, the threshold of 31.5 units described above and in \cite
{Delyfer-2011} was used so as to ensure that only genes with an
unambiguous presence be considered for addition to the network.
This approach has been successfully applied to identify proteins that
are part of regulatory cycles and which are themselves not regulated at
the level of transcription, but rather by either phosphorylation [\cite
{Guerin-et-al-2012}] or proteins in the same pathway that are more
weakly regulated.
Using this approach, we
constructed an extended network of 50 proteins from the initial network
and 68 additional proteins from bicluster 4, which could then be
connected to the network because of the addition of 192 novel proteins
that were not present in bicluster 4 (Figure~\ref{fig:Bicluster4:network}).

%f6 #&#
\begin{figure}

\includegraphics{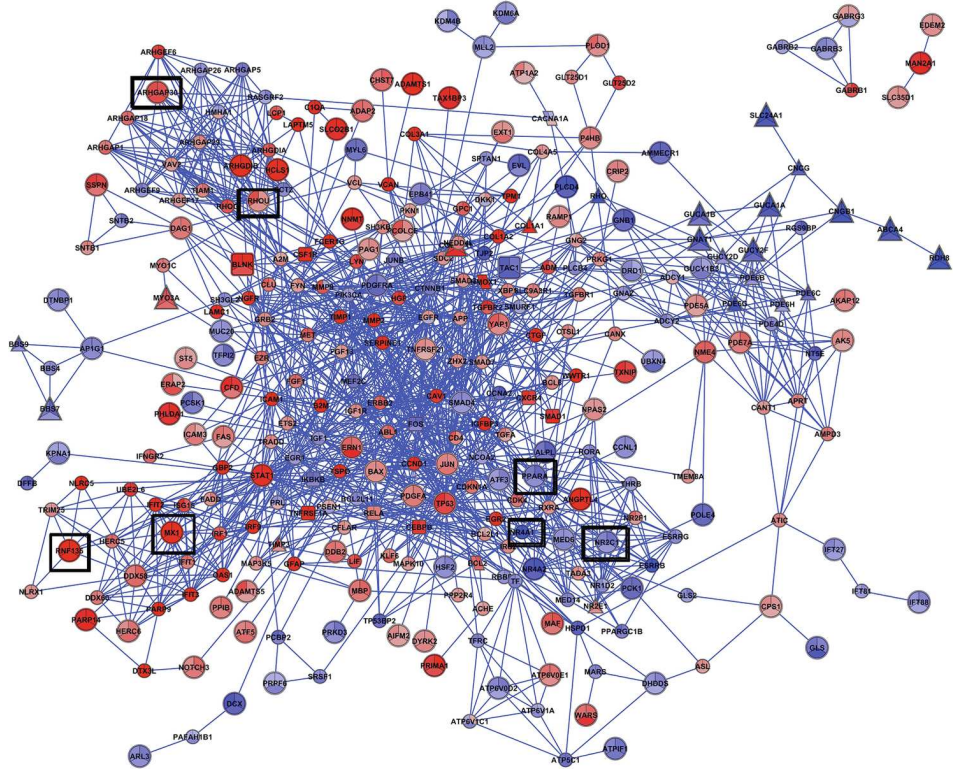}

\caption{Network map of retinal detachment transcriptome data.
Bicluster 4 from the DIC results was analyzed for protein-protein
interaction networks (PPIN) using the STRING database.
As the initial network of proteins with direct interactions was fairly small,
an extended network of 310 proteins was constructed based on 118
proteins from bicluster 4 and 192 added proteins (smaller node size in
the figure).
Relative change in biopsies for three subgroups of RD latency are
represented by different colors of nodes (proteins). Selected classes
of GO-ontologies are shown via the node shape:
triangles for ``GO:0007601 visual perception,'' parallelograms for
``GO:0008219 cell death''
and rectangles for ``GO:0006954 inflammatory response.''
Genes central to selected pathways further discussed in Section~\protect
\ref{sec:applications:RD}
are surrounded (highlighted) by black rectangles.}
\label{fig:Bicluster4:network}
\end{figure}

In the extended network, the proteins identified in bicluster~4 are
shown as large nodes, whereas the added proteins are shown as small
nodes. All nodes (proteins) are divided into three regions that
correspond to early, middle and late latency of RD. The regions are
colored according to the change of gene expression values (fold-change)
relative to the control group. The three respective fold-change values
are displayed in a blue to red color scale (saturated blue for
down-regulation stronger than 6-fold; saturated red for up-regulation
stronger than 6-fold). It is important to note that the majority of
proteins added to construct this extended network have node colors that
are similar to the color of their neighbors originally identified in
bicluster 4. This confirms that adding these genes conserves well the
overall structure of up- or down-regulated groups of proteins.
Several GO ontology features are displayed in Figure~\ref
{fig:Bicluster4:network}
according to the following shapes of the nodes:
triangles display genes with ``GO:0007601 visual perception,''
parallelograms, genes with
``GO:0008219 cell death,'' and rectangles, genes with ``GO:0006954
inflammatory response.''
No cases of multiple annotations combining any of these three terms
were observed among the 310 proteins that form this network. Genes
annotated with other functions are shown as circles.
Proteins involved in cell death and inflammation were key results in
the traditional analysis using $t$-tests [\cite{Delyfer-2011}].
In contrast, proteins with these annotations are fairly rare in DIC
bicluster 4,
and are found in separate substructures of the enriched network when
compared to the down-regulated genes annotated as being involved in
visual perception.
In fact, most other subnetworks based on DIC bicluster 4 are somehow
related to signaling, and thus reflect substantial biological and
molecular activity in specimens of RD.
One may note other relevant subnetworks, such as the one around
RHOU and ARHGAP30 (framed by rectangles at the top left part of
Figure~\ref{fig:Bicluster4:network}), which is highly enriched in GTPases,
which in turn are found at the very end of signaling pathways;
the subnetwork around MX1 and RNAF135 (framed by rectangles at the
bottom left part of Figure~\ref{fig:Bicluster4:network}), which is
enriched in up-regulated antiviral activity; or the subnetwork around
PPARA, NR4A2 and NR2C1 (framed by rectangles at the bottom right of
Figure~\ref{fig:Bicluster4:network}), which is enriched in mostly
down-regulated nuclear receptors. The surprisingly strong antiviral
activity subnetwork mentioned above may be involved in the general
acute inflammatory response; however, it has not been noted in the literature.
Alternatively, these findings may open novel perspectives for
further detailed studies to investigate the potential participation of
viral infections as risk factors for RD or as factors related to a
worse prognosis at the onset of RD.

%s6 #&#
\section{Conclusion}\label{sec:conclusions}

We have proposed a model for biclustering that incorporates biological knowledge
from the Gene Ontology (GO) project and experimental conditions (if available).
We use this knowledge to specify prior distributions that account for
the dependency structure between genes and between conditions.
Our goal was to determine whether using prior information on the genes
and the conditions would improve the biological significance of the
biclusters obtained from this method. We incorporated this prior information
by efficiently modeling mutual interactions between genes (or
conditions) with
discrete Gibbs fields. The pairwise interaction between the genes is given
by entropy similarities estimated from GO. These are embedded into
a relational graph with nodes that correspond to genes and edges to
similarities.
The graph is kept sparse by filtering out gene interactions (edges)
that arise from genes that do not share much common biological
functionality as measured by GO. In some cases, the conditions may also
be compared by building a notion of similarity between them, for
example, in the case of gene expression time courses. These
similarities can also be represented
by a corresponding relational graph. To our knowledge, the introduction
of Markov models and Gibbs fields in the context of biclustering is
new. However, this has already been attempted in the fields of
clustering and regression.

In order to estimate the biclusters, we adopted a hybrid procedure that
mixes the Metropolis--Hastings sampler with a variant of the Wang--Landau
algorithm. To efficiently sample the labels through a block Gibbs
sampling, we used an algorithm based on the Swendsen--Wang algorithm.
Experiments on simulated data showed that our model is an improvement
over other algorithms. They also showed that criteria based on the
conditional DIC and AIC may be used to guide the choice of the number
of biclusters.

The application of Gibbs-plaid biclustering to a data set created from
RD research brings several advantages and novel insights. In
comparison to previous efforts, we noted that biclustering is much more
adaptive to biological settings, which are characterized by numerous
proteins that have multiple functions and tissues or cells of interest
that make use of multiple biological processes at the same time. A
detailed inspection of the biclustering results allowed us to identify
biclusters that are associated with all major known groups of cellular
and molecular events. Adding a protein-network component to these
results revealed several previously unknown aspects of RD that lead to
the generation of new hypotheses regarding:
(i) proteins directly involved in subsequent changes in photoreceptor
cells, and
(ii) subnetworks of proteins potentially linked to these events.\vspace*{3pt}

\section*{Acknowledgments}
The authors are grateful to LeeAnn Chastain at MD Anderson Cancer
Center for editing assistance.

\begin{supplement}[id=suppA]
\stitle{Supplement to ``The Gibbs-plaid biclustering model''\\}
\slink[doi]{10.1214/15-AOAS854SUPP} %[doi,text={...}] - jei reikia
%suskaldyti doi
\sdatatype{.zip}
\sfilename{aoas854\_supp.zip}
\sdescription{A high-resolution version of the image shown
in Figure~\ref{fig:Bicluster4:network}, as well as
the complete biclustering results associated with the RD data
have been provided as supplementary material.
A proof of the convergence of the stochastic
algorithm of Section~\ref{sec:posterior} and further details on Lin's similarity (Section~\ref{sec:prior}) are also included.}
\end{supplement}

% imsref loaded by aiste.veprauskaite, 2015-08-21 12:57:52
% imsref loaded by aiste.veprauskaite, 2015-08-21 13:16:46
%
% imsref loaded by vpetrauskas, 2015-08-24 13:22:06

\printaddresses
\end{document}